# MAST-SEY: MAterial Simulation Toolkit for Secondary Electron Yield. A Monte Carlo Approach to Secondary Electron Emission Based On Complex Dielectric Functions

Maciej P. Polak[1, *] and Dane Morgan[1, †]

[1]*Department of Materials Science and Engineering,*
*University of Wisconsin-Madison, Madison, Wisconsin 53706-1595, USA*

MAST-SEY is an open-source Monte Carlo code capable of calculating secondary electron emission using input data generated entirely from first principle (density functional theory) calculations. It utilizes the complex dielectric function and Penn's theory for inelastic scattering processes, and relativistic Schrödinger theory by means of a partial-wave expansion method to govern elastic scattering. It allows the user to include explicitly calculated momentum dependence of the dielectric function, as well as to utilize first-principle density of states in secondary electron generation, which provides a more complete description of the underlying physics. In this paper we thoroughly describe the theoretical aspects of the modeling, as used in the code, and present sample results obtained for copper and aluminum.

## I. INTRODUCTION

Irradiation of a material surface with a beam of (primary) electrons results in subsequent emission of electrons from the surface. This effect is called secondary electron emission (SEE). The ratio of the number of emitted secondary electrons to the number of primary electrons is called the secondary electron yield (SEY).

SEE and SEY are important in a number of materials applications. For example, SEE is often discussed in the context of electron microscopy, as the secondary electrons play a crucial role in digital imaging in scanning electron microscopes (SEMs) [1, 2]. Therefore, detailed knowledge of the properties of the emitted secondary electrons is essential in proper interpretation of the results. SEE is also important to take into account when designing SEMs, as the materials used for the construction may influence the results through secondary emission [3].

In some applications, materials with particularly high or low secondary electron yield are often desirable. For example, electron multipliers, often used in detectors (photomultipliers) require materials with very high SEY in order to efficiently multiply the number of incident electrons via SEE [4].

On the other hand, high-power radio frequency (RF) devices need low SEY materials in order to suppress multipactor [5, 6]. Multipactor is an effect where an avalanche of electrons exponentially grows in numbers due to SEE from the device walls. This results in degradation of the device performance and may eventually cause a complete failure of the device. Since high-power RF devices are mainly used in the aerospace industry, their reliability is a key factor and, therefore, the choice of proper materials with low SEY is essential [7–9].

Monte Carlo (MC) modeling is a powerful tool in studying SEE [10–15]. It is generally much quicker and less expensive than experimental measurements, and allows for complete control of the studied system, at least within the domain of the model. MC modeling brings a deep insight into the scattering process by providing detailed information about electron trajectories and scattering inside the sample. The use of such methods can provide a consistent and relatively quick prediction of SEY for many materials. However, MC models require at least a few fundamental input quantities, namely the dielectric function (or energy loss function (ELF)), Fermi energy, and the work function. Therefore, the utility of MC SEY modeling depends greatly on the availability of these electronic structure properties. Fortunately, these properties can now be relatively easily obtained via first-principles electronic structure methods, particularly density functional theory (DFT). Thus it is valuable to establish a code that can utilize first-principles input in state-of-the-art MC SEY modeling, which is what we have done in this work. This work presents an open source code for studying secondary electron emission called the MAterial Simulation Toolkit for Secondary Electron Yield (MAST-SEY). MAST-SEY uses state-of-the-art methods for describing electron scattering. The dielectric function approach is used for inelastic scattering [16–18] and the elastic scattering cross sections are obtained with relativistic partial-wave calculations for scattering by a local central interaction potential [19]. The general foundation of the code implementation is based on Refs. [11, 20–22]. It allows for user input of the momentum dependent dielectric function to be used in the description of the inelastic scattering processes, and also provides the option to used standard extrapolation procedures [16]. The momentum dependence may be obtained, for example, from first-principles DFT calculations. First-principles based dielectric functions with momentum dependence are far more complete than typically available for materials, potentially increasing the accuracy of the inelastic scattering processes, and therefore, the overall SEY calculation. The use of the full, momentum and energy dependent dielectric function is especially important for low energy electrons, where the use and applicability of dielectric function approximations has been questioned and debated [23]. The following sections describe the theoretical background of the electron scattering within the model, the workflow of the MAST-SEY code, and application of the code on Cu and Al metals.

* mppolak@wisc.edu
† ddmorgan@wisc.edu

## II. ELECTRON SCATTERING

The total mean free path ($\lambda_t$) is used to determine the distance an electron travels between scattering events:

$$\lambda_t^{-1} = \lambda_e^{-1} + \lambda_{in}^{-1}. \quad (1)$$

where $\lambda_e^{-1}$ and $\lambda_{in}^{-1}$ are the elastic and inelastic mean free paths, respectively. These, together with the change of direction and energy of the electrons, govern the relevant electron transport within the material. Below, the inelastic and elastic scattering properties are described in detail.

### A. Inelastic Scattering Properties

The inelastic scattering properties are calculated in accordance to the theory described by Penn [16]. In this approach the inverse inelastic mean free path for an electron of energy $E$ in a material with a wavevector and frequency dependent dielectric function $\varepsilon(q,\omega)$, and Fermi energy $E_F$ is expressed by the following equation:

$$\begin{aligned}\lambda_{in}^{-1}(E) &= \int_0^E \frac{d\lambda_{in}^{-1}}{d\omega}d\omega = \int_0^E d\omega \frac{2}{\pi v^2}\int_{q_-}^{q_+}\frac{dq}{q}\text{Im}\left(\frac{-1}{\varepsilon(q,\omega)}\right)\\ &= \int_0^E d\omega \frac{(1+E'/c^2)^2}{(1+E'/(2c^2))\pi E'}\int_{q_-}^{q_+}\frac{dq}{q}\text{Im}\left(\frac{-1}{\varepsilon(q,\omega)}\right).\end{aligned} \quad (2)$$

The imaginary part of the negative inverse of the dielectric function is called the ELF. Throughout the calculations, the bottom of the conduction band serves as energy reference, therefore the Fermi energy is calculated as the energy between the bottom of the conduction band and the Fermi level. The $q$ integration limits, in a relativistic approach, are:

$$q_\pm = \sqrt{E'\left(2+\frac{E'}{c^2}\right)} \pm \sqrt{(E'-\omega)\left(2+\frac{E'-\omega}{c^2}\right)}, \quad (3)$$

where $E' = E - E_F$ and $c$ is the speed of light. A non-relativistic formulation also exists, but since it is only marginally more computationally advantageous, the relativistic approach is always used within MAST-SEY. The most common approach to obtain the momentum dependent ELF is an extrapolation of the $q=0$ curve to finite $q$ values through a proper dispersion relation:

$$\text{Im}\left(\frac{-1}{\varepsilon(q,\omega)}\right) = \frac{\omega_0}{\omega}\text{Im}\left(\frac{-1}{\varepsilon(q=0,\omega_0)}\right) \quad (4)$$

where $\omega_0$ is the $\omega_p$ fulfilling the plasmon dispersion equation:

$$\omega_q(q,\omega_p) = \omega \quad (5)$$

By default, in order to achieve optimal results, MAST-SEY expects a user-provided $q$-dependent ELF (or dielectric function) directly as an input. This approach is potentially superior to extrapolation methods because it allows the use of directly obtained $q$ dependence. In particular, such data can be obtained through density functional theory first principle calculations. MAST-SEY also allows for utilizing $q = 0$ ELF (or dielectric function), e.g. obtained experimentally, on which different $q$ extrapolation schemes, based on the Penn approach [16], be can be performed. In the following subsections we provide a summary of the approaches to obtaining the wavevector dependence of the ELF.

#### 1. Single Pole (Simple Penn) Approximation - SPA

In the Penn approach, the expression for ELF can be written as:

$$\text{Im}\left(\frac{-1}{\varepsilon(q,\omega)}\right) = \int_0^\infty g(\omega_p)\text{Im}\left(\frac{-1}{\varepsilon_L(q,\omega;\omega_p)}\right)d\omega_p, \quad (6)$$

where $\varepsilon_L$ is the dielectric function according to the Lindhard model [24] for a free electron gas, $\omega_p$ is the plasmon frequency and $g(\omega) = \frac{2}{\pi\omega}\text{Im}\left(\frac{-1}{\varepsilon(\omega)}\right)$. In the single pole approximation, the Lindhard ELF can be approximated by:

$$\text{Im}\left(\frac{-1}{\varepsilon_L(q,\omega;\omega_p)}\right) \approx \frac{\pi\omega_p^2}{2\omega_q(q,\omega_p)}\delta\left(\omega-\omega_q(q,\omega_p)\right), \quad (7)$$

where:

$$\omega_q(q,\omega_p) = \sqrt{\omega_p^2 + \frac{1}{3}(k_F(\omega_p)q)^2 + \frac{q^2}{4}}, \quad (8)$$

and $k_F(\omega_p) = \sqrt[3]{\frac{3}{4}\pi\omega_p^2}$. Integration of Eq. 6 leads to:

$$\text{Im}\left(\frac{-1}{\varepsilon(q,\omega)}\right) = \text{Im}\left(\frac{-1}{\varepsilon(\omega_0)}\right) \Big/ \left(1 + \frac{\pi q^2}{6k_F(\omega_0)}\right), \quad (9)$$

where $\omega_0$ is the positive real solution of $\omega_q(q,\omega_0) = \omega$.

#### 2. *Simplified Single Pole (Simplified Simple Penn) Approximation - SSPA*

Substitution of Eq. 8 with a simple quadratic plasmon dispersion

$$\omega_q(q,\omega_p) = \omega_p + \frac{q^2}{2} \quad (10)$$

leads to an expression for simplified single pole approximation:

$$\text{Im}\left(\frac{-1}{\varepsilon(q,\omega)}\right) = \text{Im}\left(\frac{-1}{\varepsilon(\omega_0)}\right)\frac{\omega_0}{\omega}, \quad (11)$$

where, similarly to the previous case, $\omega_0$ is the positive real solution of $\omega_q(q,\omega_0) = \omega$.

### B. Elastic Scattering Properties

The elastic scattering of electrons in a medium is a complex process and numerical models are already thoroughly described in literature [19, 25–27]. Therefore, in our MAST-SEY code we utilize the capabilities of the `elsepa` [19] code in order to obtain the energy and angle dependent elastic differential cross sections. Below we provide a brief summary of the methods used to obtain the inverse inelastic mean free paths. An extensive and detailed description of all the theoretical procedures can be found in [19], alongside with all the default numerical parameters used. MAST-SEY by default uses the parameters recommended by the authors of [19], however there is some freedom in choosing the models of nuclear charge distribution and electron distribution, as well as the electron exchange potential. The nuclear charge distribution can be chosen from point nucleus (P), uniformly charged sphere (U), Fermi distribution (F), or Helms uniform-uniform distribution (UU). The electron distribution can be chosen from Thomas-Fermi-Molire distribution (TFM), Thomas-Fermi-Dirac distribution (TFD), Dirac-Hartree-Fock-Slater distribution (DHFS), or numerical Dirac-Fock distribution (DF) read from the database files. The exchange potential can be chosen from Furness-McCarthy potential (FM), Thomas-Fermi potential (TF), Riley-Truhlar potential (RT), or no exchange potential can be used. The user can also choose to include solid-state effects and correlationpolarization potential.

The energy and angle dependent elastic differential cross sections are then integrated in order to obtain the total elastic cross section:

$$\sigma_e = \int \left(\frac{d\sigma_e}{d\Omega}\right) = 2\pi \int_0^\pi \left(\frac{d\sigma_e}{d\Omega}\right)\sin\theta d\theta \quad (12)$$

which is then subsequently used to calculate the elastic mean free path (EMFP):

$$\lambda_e = \frac{A}{N_a \rho \sigma_e} = \frac{V}{\sigma_e}, \quad (13)$$

where $N_a$ is the Avogadro number, $\rho$ is the mass density, $A$ is the atomic mass, and $V$ is the volume per one atom.

### III. MONTE CARLO PROCEDURE

The scattering processes are governed by random numbers $R_n \in [0,1)$. The process starts with an electron entering the material from vacuum. The initial energy of the electron in the material is expressed as $E = E_v + U_0$ where $U_0 = E_F + \phi$ is the inner potential, expressed as a sum of the Fermi energy $E_F$ and work function $\phi$, and $E_v$ is the energy of the incoming electron in vacuum. When crossing the boundary, the polar angle of the incoming electron with respect to the surface normal changes according to $\alpha = \text{asin}(\sin(\alpha_v)\sqrt{E_v/E})$, where $\alpha$ is the initial polar angle inside the material and $\alpha_v$ is the initial polar angle in the vacuum before entering the material. Next, the distance $s$ an electron travels between scattering events is determined, based on the total mean free path (Eq. 1):

$$s = -\lambda_t \ln R_1. \quad (14)$$

The type of scattering event is determined by a random number. It is elastic if the following is true:

$$\frac{\lambda_e^{-1}}{\lambda_t^{-1}} > R_2, \quad (15)$$

otherwise, the electron is scattered inelastically.

### A. Elastic electron scattering

An elastic collision results only in the change of the angles at which the incident electron travels. A change in the polar angle $\theta$ is calculated according to:

$$\int_0^\theta \left(\frac{d\sigma_e}{d\Omega}\right)\sin\theta' d\theta' = R_3 \int_0^\pi \left(\frac{d\sigma_e}{d\Omega}\right)\sin\theta' d\theta'. \quad (16)$$

A change in the azimuthal angle $\phi$ is simply expressed by:

$$\phi = 2\pi R_4 \quad (17)$$

### B. Inelastic electron scattering

An inelastic collision results in a loss of energy by the incident electron, an associated change in the angles at which the electron travels and a generation of a new secondary electron. The amount of energy lost $\Delta E$ is calculated with the use of:

$$\int_0^{\Delta E} \frac{d\lambda_{in}^{-1}}{d(\Delta E')}d(\Delta E') = R_5 \int_0^{E-E_F} \frac{d\lambda_{in}^{-1}}{d(\Delta E')}d(\Delta E'). \quad (18)$$

The amount of the energy lost governs the change in the polar angle ($\theta$):

$$\int_0^\theta \frac{d^2\lambda_{in}^{-1}}{d\Omega d(\Delta E)} \sin\theta' d\theta' = R_6 \int_0^\pi \frac{d^2\lambda_{in}^{-1}}{d\Omega d(\Delta E)} \sin\theta' d\theta', \quad (19)$$

where the integrand can be expressed as:

$$\frac{d^2\lambda_{in}^{-1}}{d\Omega d(\Delta E)} = \frac{1}{\pi^2 E} \mathrm{Im}\left(\frac{-1}{\epsilon(q,\omega)}\right) \frac{1}{q^2} \sqrt{E(E - \Delta E)}, \quad (20)$$

with

$$\frac{q^2}{2} = 2E - \Delta E - 2\sqrt{E(E-\Delta E)}\cos\theta'. \quad (21)$$

Alternatively, an approximate rudimentary formula based on classical inelastic collision may be used $\sin\theta = \sqrt{\Delta E/E}$. This significantly improves calculation speed, but at a cost of accuracy.

The change in the inelastic scattering azimuthal angle $\phi$ is similar to the case of elastic scattering:

$$\phi = 2\pi R_7 \quad (22)$$

Each inelastic scattering event produces a secondary electron. MAST-SEY allows the user to choose how the secondary electron is generated from the options described next. In the simplest case the secondary electron is generated from the Fermi sea, therefore it has an initial energy of $E_s = \Delta E + E_F$. Alternatively it may be generated from the valence band with energy $E_s = \Delta E + E_V$, where the valence energy ($E_V$) is obtained using the joint density of states (JDOS) $JD(\Delta E, E_V) = D(\Delta E + E_V)D(E_V)$, where $D(x)$ is the density of states, and another random number according to:

$$\int_0^{E_V} JD(\Delta E, E)dE = R_8 \int_0^{E_F} JD(\Delta E, E)dE, \quad (23)$$

assuming the bottom of the valence DOS is at $E = 0$. The DOS may be either derived from an ideal parabolic band dispersion (i.e. the simple case of a free electron gas), or may be provided by the user, as a result of a more advanced approach e.g. first principle DFT.

Optionally, if the loss of energy $\Delta E$ is high enough ($\Delta E > E_b$), the secondary electron may be generated with energy equal to $E_s = \Delta E - E_b$ from a bound state of energy $E_b$.

The direction of the generated secondary follows $\theta_s = \arcsin(\cos\theta)$ and $\phi_s = \pi + \phi$.

The secondary electron generated as an effect of inelastic scattering is then treated in the same way as a primary electron.

### C. Secondary Electron Emission

An electron, following a series of scattering events, may reach back to the surface of the material. If that happens it may escape and contribute to the secondary electron yield or be reflected back. Whether it does escape or not, is governed by the transmission $T(E, \beta)$ function, obtained from a solution to the Schrödinger equation for a step potential:

$$T(E,\beta) = \begin{cases} \frac{4\sqrt{1-\frac{U_0}{E\cos^2\beta}}}{\left(1+\sqrt{1-\frac{U_0}{E\cos^2\beta}}\right)^2}, & \text{if } E\cos^2\beta > U_0 \\ 0, & \text{if } E\cos^2\beta \leq U_0, \end{cases} \quad (24)$$

here $\beta$ is the angle at which the electron approaches the surface and $U_0 = E_F + \phi$ is the inner potential, expressed as a sum of the Fermi energy $E_F$ and work function $\phi$. The emitted electron has an energy $E_{em} = E - U_0$ and is emitted with a changed polar angle $\beta_{em} = \arcsin(\sqrt{E}\sin\beta/\sqrt{E-U_0})$.

### D. Alloys

MAST-SEY allows the user to perform calculations for alloys as well. In this case, the inelastic scattering processes are calculated directly from the alloy's dielectric function (as described above in Sec. II A. The elastic scattering is treated with a method similar to the virtual crystal approximation (VCA) [28], i.e. $\frac{d\sigma_e}{d\Omega}$ (see Eq. 12) is calculated for all alloying elements and then averaged according to the composition. The errors introduced by this approximate approach have not been thoroughly investigated and more work is needed to determine its accuracy. However, similar to the VCA, it is expected to perform best for alloys composed of similar elements.

## IV. WORKFLOW OF MAST-SEY

In order to obtain the secondary electron yield from dielectric data, the code is executed in two stages, a preparation step and an execution step, described below.

### A. Preparation step

In the first step, the executable is ran with a `prepare` option. The dielectric function, given in an input file either as a dielectric function directly (`eps.in`) or as the energy loss function (`elf.in`), is used to calculate energy dependent cumulative integrals of differential inelastic cross sections (Eq. 2). For this calculation, a momentum dependent energy loss function is necessary. The flag `-qdep` allows to specify how the momentum ($q$) dependence of ELF is to be obtained, where the user can choose from `SPA`, `SSPA` and `CUSTOM`. Section II A describes the first two keywords - extrapolation approaches available in MAST-SEY. If the `CUSTOM` keyword is used, the momentum and energy dependent ELF will be read from the `elf.in` file and will not undergo any further treatment. This option is used in this work when DFT obtained ELF is considered. At the same time, during the first step, energy dependent cumulative integral of differential elastic cross sections are generated (Eq. 13). The flag `-elastic` allows the user to specify the formalism for generation of the elastic mean free paths via the `elsepa` package. The details can be found in Sec. II B. The preparation step is performed for a chosen energy range not exceeding the energy range of the provided dielectric function.

The material constants needed for the mean free path calculations are provided in the mandatory `material.in` file. It contains values such as the atomic number (or numbers and their compositions, in the case of alloys), volume per atom, Fermi energy and work function. The results are saved in files `inelastic.in` and `elastic.in`, respectively. Additionally, an `mfp.plot` file is created, allowing for easy plotting of the resultant mean free paths.

MAST-SEY was intentionally designed to perform the preparation step separately from the MC simulation. The preparation step requires significant computational time and the created files can be reused for different MC simulation parameters. Therefore it is preferable to perform this step only once. More detailed information on the preparation step, as well as a short tutorial with examples, can be found in the manual available alongside the source files for the code. Due to the nature of the code, changes may be introduced over time, the entire history of which will be available in the code manual. Quick access to the list of currently available options is available executing the code with the help flag `-h`.

### B. Simulation step

The second step is performing the actual MC simulation with the use of the previously prepared files. Here, values such as the incident energy (`-e` flag) and number of incident electrons (`-m`) are entered. The angle of incidence of incoming electrons, normal to the surface by default, can be specified with the optional `-pe` flag. The code will then read the previously prepared `*.in` files and perform Monte Carlo simulations. Five SEY related values are generated as an output: total SEY, true SEY, backscattering, diffused primaries and elastically backscattered electrons, they are all ratios i.e. represent number of emitted electrons per one incident electron. True SEY and backscattering correspond to electrons emitted with energy $\leq 50$ eV and $> 50$ eV, respectively. All of the electrons emitted from the entire hemisphere above the surface are considered. By default, these values are the only output of the code, in order to preserve memory consumption. However, with the use of additional flags such as `-coord` or `-distr`, for example, additional information can be obtained, such as the coordinates of the trajectories of each of the electrons or energy and angle distribution of the emitted secondaries. In order to obtain a SEY curve (as a function of the energy of the incident electrons), the simulation step has to be run separately for each desired incident energy. If multiple CPU cores are available, each subsequent execution will occupy a different core. Since the code does not require significant amount or memory, this is the most efficient mode of parallelization and conveniently does not require compilation with MPI libraries. Parallelization tools such as GNU Parallel [29] may be used as well to simplify the parallelization process. Again, a comprehensive and detailed description of flags and their proper use is available in the code manual, with a list of available options quickly accessible by executing the code with the help flag `-h`.

## V. EXAMPLE RESULTS

In order to present the capabilities of the code as well as verify its results, we chose two widely studied and distinct materials: copper, a non free-electron like material, and aluminum, a free-electron like material. They differ significantly in their dielectric functions, and, therefore, their energy loss functions. Copper, generally, has an ELF of a relatively low magnitude spread out into a broad spectrum. The ELF of aluminium, on the other hand, has a single sharp plasmon peak in the low energy range. These features are clearly visible in Figs. 1 and 2. It is expected that the different ELFs will have significantly different character of the $q$-dependence. This $q$-dependence is directly factored in the double differential cross section for electron inelastic scattering (the expression under the energy integral in Eq. 2) through the integral with respect to $q$, which is the last term in Eq. 2. Since inelastic scattering is the main energy loss mechanism, it is expected to have a significant impact on the SEY. It has been shown previously that different treatments of the momentum dependence may result in a dramatically different SEY. For example, in Ref. [23], the use of a more accurate full Penn approximation (FPA) compared to the less accurate simple Penn approximation SPA, reduce the resulting SEY by over 50%. The same comparison for Cu has shown only a small change in the results when FPA is used instead of SPA. This proves the choice of different extrapolations of the ELF into the momentum space may significantly impact the final result. In this work, due to the possibility of using a user provided energy and momentum dependent ELF, we predict the SEY without the need for extrapolation approximations such as the SPA or FPA. In the following sections we compare SEY curves obtained with the use of MAST-SEY for two different approaches. The first approach included only the most essential input, i.e. the optical ($q = 0$) dielectric function (in the form of the ELF), the Fermi energy ($E_F$) and the work function ($\phi$). The secondaries were generated from a free electron DOS, and the $q$-dependence was obtained using SPA. The second approach utilized the full suite of capabilities of MAST-SET and included the information available only through the explicit use of first principle calculations. A DFT calculated momentum and energy dependent ELF was employed, with the secondary electrons generated with the use of DFT obtained density of states.

The Fermi energy was calculated from the DFT obtained density of states and the work functions from [30] (weighted average value) were used, where they have also been calculated with first principle DFT methods.

### A. Simulation of SEY of Copper and Aluminium

To demonstrate the capabilities of the code, we used copper and aluminium as examples. Copper is a system that is probably the most often studied both experimentally and theoretically and frequently used as a benchmark. It is also a good representative of a non-free electron like metal. This is reflected in a highly non-parabolic DOS as well as a broad ELF. Aluminium, on the other hand, being a free-electron like metal, exhibits an almost parabolic DOS and an ELF dominated by a single sharp peak. All figures and information provided below have been obtained directly as a result of MAST-SEY runs.

Figures 1 and 2 show momentum and energy dependent ELFs for Cu and Al calculated with two methods for q-space extrapolation: SPA - Eq. 9, and DFT - directly calculated from first principles. The broad ELF of Cu and sharp peak in ELF of Al are clearly visible.

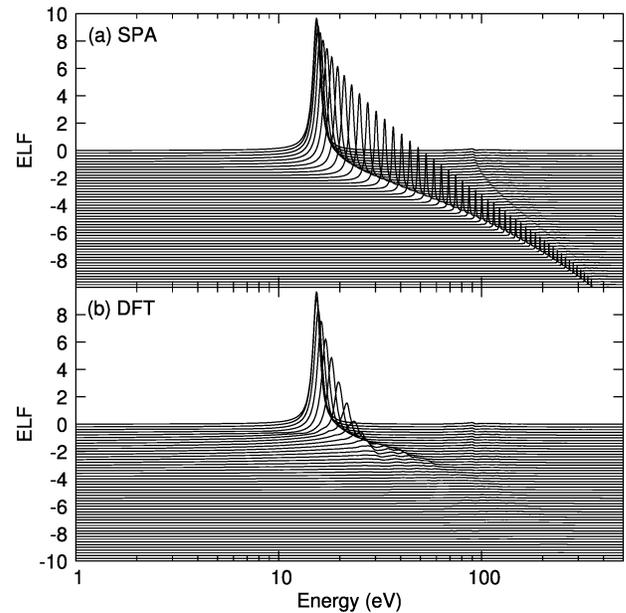

FIG. 2. Energy and momentum dependence of the energy loss function of aluminium. a) single pole approximation (SPA), b) DFT calculations. Lines are offset by -0.02 and represent increasing q-values ranging from $q = 0.001$ Bohr$^{-1}$ to $q = 5$ Bohr$^{-1}$ in increments of $q = 0.1$ Bohr$^{-1}$.

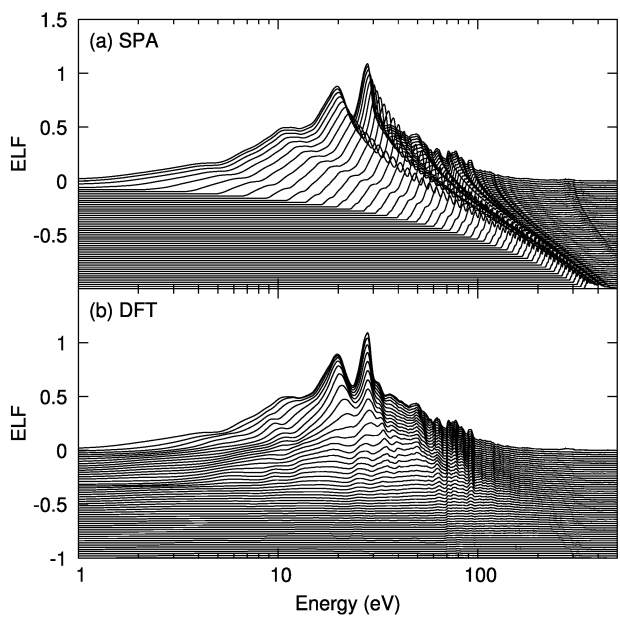

FIG. 1. Energy and momentum dependence of the energy loss function of copper. a) single pole approximation (SPA), b) DFT calculations. Lines are offset by -0.005 and represent increasing q-values ranging from $q = 0.001$ Bohr$^{-1}$ to $q = 5$ Bohr$^{-1}$ in increments of $q = 0.1$ Bohr$^{-1}$.

A very different behavior is observed between the two approaches. The DFT calculations exhibit a much faster decay of the ELF as a function of momentum. Additionally, in SPA the entire ELF is treated with the same dispersion relation. In DFT on the other hand, different features of ELF change as a function of momentum differently, both in terms of magnitude as in relative position on the energy scale. This behavior is expected of a proper momentum dependent ELF, as different features (peaks) may have different physical origin and can be attributed to different physical properties [31].

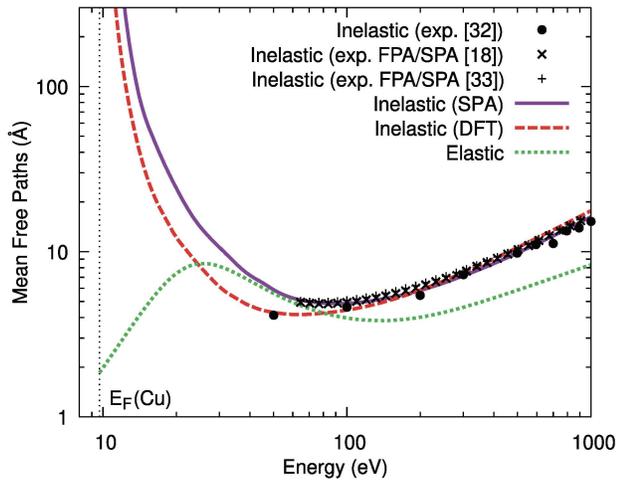

FIG. 3. Mean free paths of copper. Inelastic mean free paths were obtained with single-pole approximation (blue solid curve) and first-principle DFT (red dashed curve). Elastic mean free path (green dotted curve) was obtained with Fermi distribution of nuclear charge, Dirac-Fock model for electron distribution and Furness-McCarthy exchange potential. Vertical dotted line marks the Fermi energy. Points (solid circles) correspond to elastic-peak electron spectroscopy measurements from [32], while crosses (overlapping) correspond to results of optical measurements combined with full and simplified Penn approximations [18, 33].

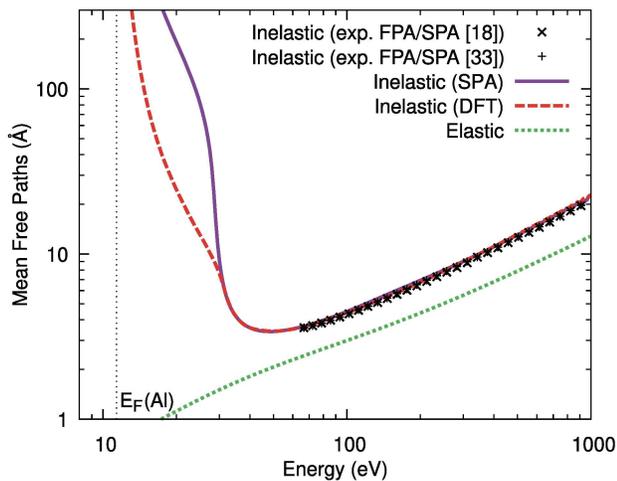

FIG. 4. Mean free paths of aluminum. Inelastic mean free paths were obtained with single-pole approximation (blue solid curve) and first-principle DFT (red dashed curve). Elastic mean free path (green dotted curve) was obtained with Fermi distribution of nuclear charge, Dirac-Fock model for electron distribution and Furness-McCarthy exchange potential. Vertical dotted line marks the Fermi energy. Points (overlapping) correspond to results of optical measurements combined with full and simplified Penn approximations [18, 33].

Figures 3 and 4 show mean free paths of copper and aluminium, respectively. The shapes of the curves for the two materials are distinctly different due to the very different characters of ELF. The influence of the different approaches to the q-space extrapolation (SPA vs. DFT), also has a different effect on the inelastic mean free path (IMFP) for the two materials. For copper (Fig. 3) the shape of the curves are similar, with the minimum for DFT slightly shifted to lower energies and shorter IMFPs. The low energy part of DFT obtained IMFPs have slightly lower values. In the case of Al, the maxima and high energy part of the IMFPs are virtually the same, however, there is a very pronounced difference in the shape and values of IMFPs for the low energy part of the curve, where the DFT obtained IMFP has a significantly lower value. Most of the IMFPs available in literature are based on optical measurements and analytical $q$-dependence [18, 33]. The IMFPs calculated here with the use of SPA closely follow curves obtained with the use of optical measurements and Penn approximations. This validates the accuracy of the $q = 0$ energy loss function obtained and used here. Direct experimental measurements of IMFPs are difficult, therefore are very scarce. Such data is available for copper [32], and is plotted in Fig. 3 with solid circles. The agreement with the DFT obtained IMFP is very good and the DFT based IMFPs match the experimental data in low energy regions, where Penn approximations do not. This suggests that the DFT-based approach is superior over the Penn approximations. It is especially important that the DFT is more accurate in the low energy range of IMFPs as this energy range is crucial for proper description of SEY, as will be explained later in the text.

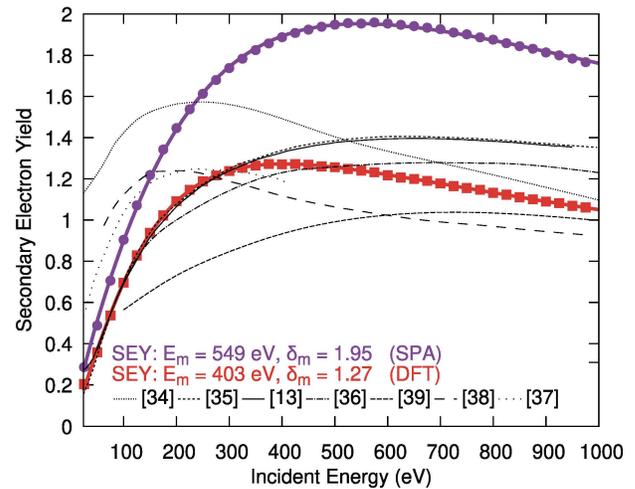

FIG. 5. Secondary electron yield (SEY) of copper. Blue and red colors correspond to single-pole approximation and DFT q-dependence, respectively. Colored lines are best fits to Eq. 25. Black lines correspond to various experimental measurements [13, 34–39]

Figures 5 and 6 show the total secondary electron yield for Cu and Al, respectively. Red points correspond to IMFPs obtained with SPA and blue points to IMFPs obtained directly from DFT. Continuous colored lines are

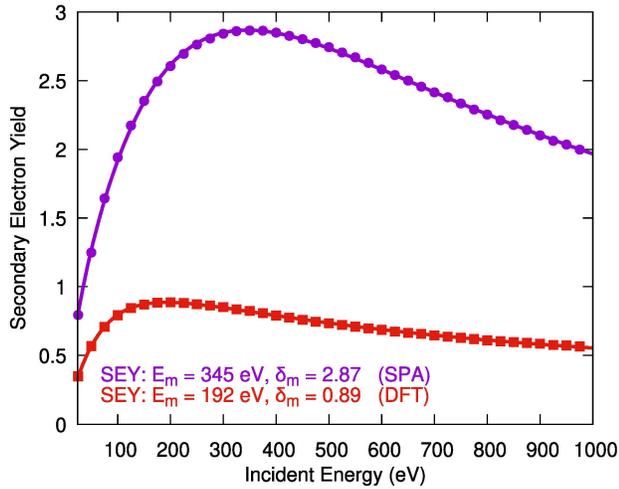

FIG. 6. Secondary electron yield (SEY) of aluminium. Blue and red colors correspond to single-pole approximation and DFT q-dependence, respectively. Lines are best fits to Eq. 25.

fits to the model equation [40]:

$$\delta = \delta_m (n_2 - n_1) \left( (n_2 - 1)\left(\frac{E}{E_m}\right)^{n_1 - 1} - (n_1 - 1)\left(\frac{E}{E_m}\right)^{n_2 - 1} \right)^{-1}, \quad (25)$$

where $n_1$, $n_2$, $\delta_m$, and $E_m$ are fitting parameters. The fits were performed in order to obtain the maximum value of SEY ($\delta_m$) and the energy at which the said maximum is observed ($E_m$) in a consistent way for both materials. The values obtained from the fit are presented on the plot. Apart from the obvious differences in SEY between the two systems, the effect of different treatment of the q-space extrapolations is different as well. In copper, the maximum SEY is reduced by around 35%. For Al, the reduction is much more dramatic, with a 69% reduction. This result, together with the observations for how the IMFPs themselves differ (Figs. 3 and 4) and the inelastic scattering energy distribution (Fig. 7) suggest, that the proper description of the low energy part of IMFP is crucial for SEY.

Figure 7 shows a distribution of the energies of the electrons at which inelastic scattering effects occur, for both Al and Cu, as calculated with the DFT obtained q-dependence of ELF. As can be clearly seen, the majority of the inelastic scattering takes place at low energies. Electrons with energy close to the Fermi energy, with the mean around 17.1 eV for Al and around 16.1 eV for Cu (for incident energy of 275 eV which is close to the energies corresponding to maximum SEY). This allows for a better understanding of the differences in SEY observed between the DFT and SPA approaches (red and blue curves on Figs. 3 and 4). The majority of the differences between the IMFPs obtained with these two approaches is visible at low energies, close to those at which the inelastic scattering events most often occur.

A very similar analysis has been performed in [23], where the SPA approach has been compared to FPA and the conclusion drawn for Al are the same. However, for copper, the FPA approach yielded relatively similar results to SPA. This suggests that even FPA, which is a relatively advanced method of ELF extrapolation, does not capture all the intricacies of the momentum dependence of the ELF.

Although different levels of complexity in different q-space extrapolation of ELF are possible (Sec. II A), direct calculation from first principles undoubtedly provides the most complete picture. Therefore we propose the use of DFT calculated energy and momentum dependent ELF as the method of choice for calculating IMFP for SEY. MAST-SEY is the first openly available code that allows for using that approach.

The main purpose of this paper is to describe the new code, its capabilities, and the theory behind it. However, a direct comparison to experimental results would help validate the approach. Unfortunately, a definitive comparison to experiment is very difficult in this case, as the SEY values and curves reported in literature are very inconsistent [15, 41]. These variations are often attributed to the surface of the sample, its cleanliness and oxidation, with maximum SEY significantly decreasing for samples submitted to surface cleaning treatments or bake-out [42].

A number or experimental results for copper seem to be relatively consistent between each other, considering the issues described above. Figure 5 contains a comparison of the results obtained here to these experiments. The agreement is clearly much better when using the DFT input, further supporting the value of this new capability. The maximum value of SEY and the first crossover point, arguably two most important material parameters governing multipactor, are in good agreement with most of the available experimental data. The shape of the curve, however, seems to deviate from the experiments. It is difficult to know the source of this discrepancy at this time due to the many approximations in this type of modeling and challenges in extracting SEY data from clean samples that match what is being modeled. Detailed comparisons with experiments will be a focus of our future work and we hope that our development of MAST-SEY will encourage others to further explore this issues as well. Experimental data reported for aluminum seems to be less consistent (not shown here as it varied widely and no reasonable quantitative comparison is possible), but the more recent ones suggest a maximum SEY for clean aluminum samples to be smaller than 1.8 [43], or even as low as 0.95 [42] value is much closer to the 0.89 obtained here with DFT than the value of 2.87 obtained with SPA, which value is clearly overestimated. All these conclusions indicate that the use of the new DFT based approach available in MAST-SEY significantly improves the result compared to experiment over the SPA approach, but that more work is needed to enable successful quantitative comparison between SEY experiments and modeling.

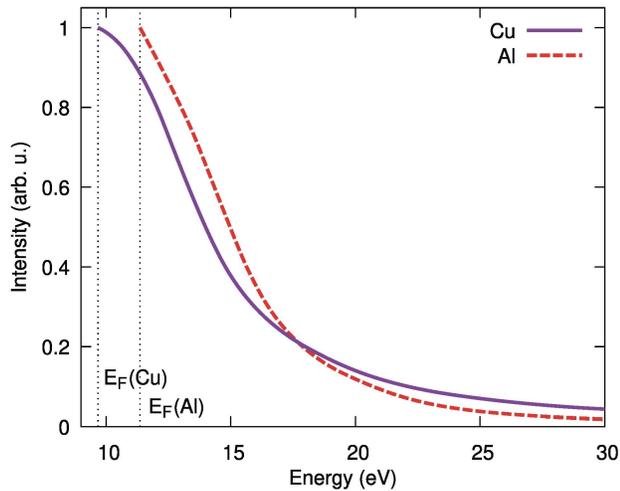

FIG. 7. Energies at which inelastic scattering events occur, for electron incident energy of 275 eV. Horizontal lines mark Fermi energies.

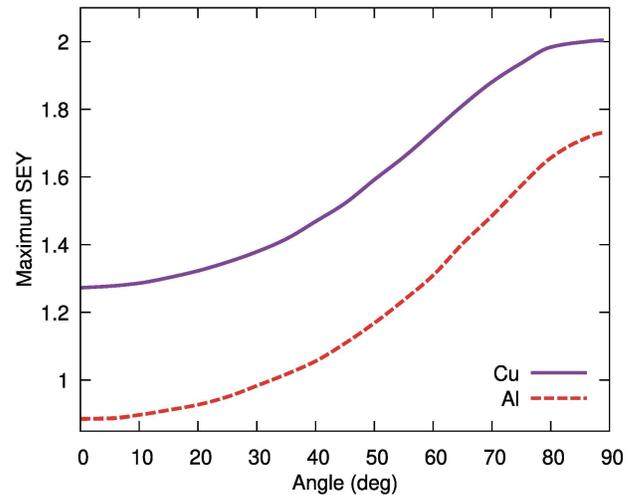

FIG. 9. Maximum SEY as a function of the angle of incidence of incoming electrons.

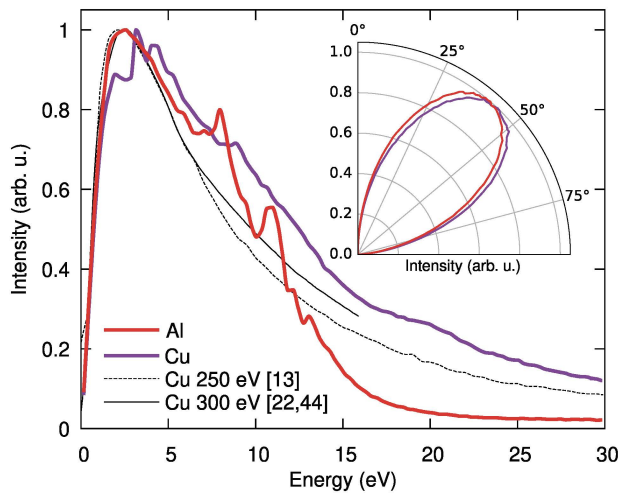

FIG. 8. Energy and polar angle (inset) distribution of the emitted secondary electrons for aluminum and copper, for incident energy of 300 eV. Thin black lines correspond to literature data [13, 22, 44].

Fig. 8 shows the energy and angle distribution (inset on Fig. 8) of the emitted electrons for both Al and Cu (for incident energy of 275 eV). It can be clearly seen that even though the materials are quite different, the energy at which the largest amount of secondary electrons are emitted is similar for both systems and occurs at around $2.5-3.0$ eV. The emission angle (polar angle with respect to the surface normal) is also very close and follows the Lambert's cosine law [45]. It is important to point out that the energy distribution curves calculated within this work exhibit some fine structure of their shape, contrary to the more smooth curves obtained from previous research and experimentally. This fine structure can be attributed directly to the fact that we use a more accurate and detailed DFT-obtained density of states in the process of secondary electron generation rather than an ideal free-electron gas DOS or a single value of energy for Fermi excitation. The distribution of energy of the secondaries within the material is governed by the joint density of states, therefore the energy of the electrons reaching the surface and eventually escaping is heavily influenced by the density of states as well. We believe that this fine structure is correct and would be present in experimental data if it had adequate resolution and no other sources of noise. However, the experimental data and theoretically fitted models describe a real sample with local non-uniformities such as surface roughness (and associated changes in the work functions), defects, impurities etc. which all combined may obscure the DOS effects. Conversely, experimental measurements that show this fine structure may indicate a sample of high quality (i.e., low impurities, defects, surface roughness, etc.). For copper, theoretical data [22, 44] (solid black line), as well as experimental results [13] (dashed black line) of energy distribution are available. The curves were normalized by magnitude to allow for direct comparison, which results in a satisfactory, although not perfect, agreement. The agreement might be improved by normalizing by area over the entire energy range, however the reference data is not available for that comparison.

Finally, to demonstrate the capability of the code to calculate SEY for different incident electron beam angles, the incident beam angle dependent SEY of Cu was calculated and shown in Fig. 9. The curve exhibits the expected dependence, with SEY increasing with the increase of the angle and reaching maximum for high values of angle. This behavior is in agreement with previously reported data for other systems [46–48].

B. Details of the DFT calculations

The DFT calculations were performed with the `exciting` [49] code. For the geometry optimization, PBEsol [50] functional was used. The dielectric function was within the time-dependent DFT (TDDFT) approach [51]. The adiabatic LDA (ALDA) was used. A $30 \times 30 \times 30$ k-point mesh, with the plane wave basis limited by the cutoff $R_{Kmax} = 9$ was chosen as a consequence of convergence studies. The momentum dependence of the energy loss function was calculated in increasing increments, starting with an increment $q = 0.01$ Bohr$^{-1}$ between 0 and 5 Bohr$^{-1}$, $q = 0.05$ Bohr$^{-1}$ between 5 and 10 Bohr$^{-1}$, and $q = 0.1$ $Bohr^{-1}$ between 10 and 20 Bohr$^{-1}$.

ACKNOWLEDGMENTS

The authors acknowledge the support from the AFOSR MURI Grant No. FA9550-18-1-0062. This work used the Extreme Science and Engineering Discovery Environment (XSEDE), which is supported by National Science Foundation grant number TG-DMR090023. The authors thank I. Matanovic, R. Gutierrez and R. Johnson for fruitful discussion.

C. Contributions

M. P. P. performed the calculations and prepared/analyzed the results, D. M. guided and supervised the research. Interpretation of results and writing of the manuscript was done by both M. P. P. and D. M.. Both authors read, revised, and approved the final manuscript.

D. Competing Interest

The authors declare no competing interest.

E. Data Availability

The code alongside example input files and detailed manual are available on github at `https://github.com/uw-cmg/MAST-SEY`.

[1] Seiler, H. Secondary electron emission in the scanning electron microscope. *Journal of Applied Physics* **54**, R1–R18 (1983).
[2] Thiel, B. L. & Toth, M. Secondary electron contrast in low-vacuumenvironmental scanning electron microscopy of dielectrics. *Journal of Applied Physics* **97**, 051101 (2005).
[3] Goldstein, J. et al. *Scanning Electron Microscopy and X-ray Microanalysis ISBN: 0306472929*, vol. XIX (2003).
[4] Tao, S., Chan, H. & van der Graaf, H. Secondary electron emission materials for transmission dynodes in novel photomultipliers: a review. *Materials* **9** (2016).
[5] Kishek, R. A., Lau, Y. Y., Ang, L. K., Valfells, A. & Gilgenbach, R. M. Multipactor discharge on metals and dielectrics: Historical review and recent theories. *Physics of Plasmas* **5**, 2120–2126 (1998).
[6] Vaughan, J. R. M. Multipactor. *IEEE Transactions on Electron Devices* **35**, 1172–1180 (1988).
[7] Ruiz, A. et al. UHV reactive evaporation growth of titanium nitride thin films, looking for multipactor effect suppression in space applications. *Vacuum* **81**, 1493 – 1497 (2007).
[8] Wolk, D. et al. AO-4025 ITT ESA - Surface treatments and coatings for reduction of multipactor and Passive InterModulation (PIM) effect in RF components (2003).
[9] Wolk, D. et al. Coatings on Mg Alloys for Reduction of Multipactor Effects in RF Components (2005).
[10] Joy, D. Monte Carlo modeling for electron microscopy and microanalysis. *Microscopy Research and Technique* **35**, 413–413 (1996).
[11] Ding, Z.-J. & Shimizu, R. A monte carlo modeling of electron interaction with solids including cascade secondary electron production. *Scanning* **18**, 92–113 (1996).
[12] Ridzel, O. Y., Astaauskas, V. & Werner, W. S. Low energy (1-100 eV) electron inelastic mean free path (IMFP) values determined from analysis of secondary electron yields (SEY) in the incident energy range of 0.1-10 keV. *Journal of Electron Spectroscopy and Related Phenomena* (2019).
[13] Azzolini, M. et al. Secondary electron emission and yield spectra of metals from Monte Carlo simulations and experiments. *Journal of Physics: Condensed Matter* **31**, 055901 (2018).
[14] Dapor, M. Secondary electron emission yield calculation performed using two different monte carlo strategies. *Nuclear Instruments and Methods in Physics Research Section B: Beam Interactions with Materials and Atoms* **269**, 1668 – 1671 (2011). Computer Simulations of Radiation Effects in Solids.
[15] Lin, Y. & Joy, D. C. A new examination of secondary electron yield data. *Surface and Interface Analysis* **37**, 895–900 (2005).
[16] Penn, D. R. Electron mean-free-path calculations using a model dielectric function. *Physical Review B* **35**, 482–486 (1987).
[17] Shinotsuka, H., Tanuma, S., Powell, C. J. & Penn, D. R. Calculations of electron stopping powers for 41 elemental solids over the 50ev to 30kev range with the full Penn algorithm. *Nuclear Instruments and Methods in Physics Research Section B: Beam Interactions with Materials and Atoms* **270**, 75–92 (2012).
[18] Shinotsuka, H., Tanuma, S., Powell, C. J. & Penn, D. R. Calculations of electron inelastic mean free paths. X. Data for 41 elemental solids over the 50 eV to 200 keV range with the relativistic full Penn algorithm. *Surface and Interface Analysis* **47**, 871–888 (2015).

[19] Salvat, F., Jablonski, A. & Powell, C. J. ELSEPA-Dirac partial-wave calculation of elastic scattering of electrons and positrons by atoms, positive ions and molecules. *Computer Physics Communications* **165**, 157 – 190 (2005).
[20] Ding, Z. J., Tang, X. D. & Shimizu, R. Monte carlo study of secondary electron emission. *Journal of Applied Physics* **89**, 718–726 (2001).
[21] Shimizu, R. & Ze-Jun, D. Monte carlo modelling of electron-solid interactions. *Reports on Progress in Physics* **55**, 487–531 (1992).
[22] Dapor, M. *Transport of energetic electrons in solids: computer simulation with applications to materials analysis and characterization*, vol. 271 of *Springer Tracts in Modern Physics* (Springer, 2020), 3 edn.
[23] Mao, S. F., Li, Y. G., Zeng, R. G. & Ding, Z. J. Electron inelastic scattering and secondary electron emission calculated without the single pole approximation. *Journal of Applied Physics* **104**, 114907 (2008).
[24] Lindhard, J. On the properties of a gas of charged particles. *Kgl. Danske Videnskab. Selskab Mat.-fys. Medd.* **28** (1954).
[25] Yates, A. A program for calculating relativistic elastic electron-atom collision data. *Computer Physics Communications* **2**, 175 – 179 (1971).
[26] Salvat, F. & Mayol, R. Elastic scattering of electrons and positrons by atoms. schrödinger and dirac partial wave analysis. *Computer Physics Communications* **74**, 358 – 374 (1993).
[27] Salvat, F. & Fernndez-Varea, J. M. radial: A Fortran subroutine package for the solution of the radial Schrdinger and Dirac wave equations. *Computer Physics Communications* **240**, 165–177 (2019).
[28] Bellaiche, L. & Vanderbilt, D. Virtual crystal approximation revisited: Application to dielectric and piezoelectric properties of perovskites. *Phys. Rev. B* **61**, 7877–7882 (2000).
[29] Tange, O. *GNU Parallel 2018* (Ole Tange, 2018). URL https://doi.org/10.5281/zenodo.1146014.
[30] Tran, R. et al. Anisotropic work function of elemental crystals. *Surface Science* **687**, 48 – 55 (2019).
[31] Alkauskas, A., Schneider, S., Sagmeister, S., Ambrosch-Draxl, C. & Hbert, C. Theoretical analysis of the momentum-dependent loss function of bulk Ag. *Ultramicroscopy* **110**, 1081 – 1086 (2010).
[32] Tanuma, S. et al. Experimental determination of electron inelastic mean free paths in 13 elemental solids in the 50 to 5000 ev energy range by elastic-peak electron spectroscopy. *Surface and Interface Analysis* **37**, 833–845 (2005). URL https://onlinelibrary.wiley.com/doi/abs/10.1002/sia.2102.
[33] Tanuma, S., Powell, C. J. & Penn, D. R. Calculations of electron inelastic mean free paths. ix. data for 41 elemental solids over the 50 ev to 30 kev range. *Surface and Interface Analysis* **43**, 689–713 (2011). URL https://onlinelibrary.wiley.com/doi/abs/10.1002/sia.3522.
[34] Septier, A. & Belgaroui, M. Secondary electron emission yields of carbon coated copper and niobium real surfaces. *IEEE transactions on electrical insulation* 725–728 (1985).
[35] Cimino, R. et al. Detailed investigation of the low energy secondary electron yield of technical cu and its relevance for the lhc. *Phys. Rev. ST Accel. Beams* **18**, 051002 (2015).
[36] Bruining, H. & De Boer, J. Secondary electron emission: Part i. secondary electron emission of metals. *Physica* **5**, 17–30 (1938).
[37] Larciprete, R., Grosso, D. R., Commisso, M., Flammini, R. & Cimino, R. Secondary electron yield of cu technical surfaces: Dependence on electron irradiation. *Phys. Rev. ST Accel. Beams* **16**, 011002 (2013).
[38] Valizadeh, R. et al. Low secondary electron yield engineered surface for electron cloud mitigation. *Applied Physics Letters* **105**, 231605 (2014). URL https://doi.org/10.1063/1.4902993.
[39] Bronshtein, I. & Fraiman, B. S. Vtorichnaya elektronnaya emissiya.(secondary electron emission). (1969).
[40] Clerc, S., Dennison, J. R., Hoffmann, R. & Abbott, J. On the computation of secondary electron emission models. *IEEE Transactions on Plasma Science* **34**, 2219–2225 (2006).
[41] Joy, D. C. A database on electron-solid interactions. *Scanning* **17**, 270–275 (1995).
[42] Baglin, V. et al. The secondary electron yield of technical materials and its variation with surface treatments. In *7th European Particle Accelerator Conference (EPAC 2000)*, 217–221 (2000).
[43] Le Pimpec, F., Kirby, R. E., King, F. K. & Pivi, M. Electron conditioning of technical aluminium surfaces: Effect on the secondary electron yield. *Journal of Vacuum Science & Technology A* **23**, 1610–1618 (2005).
[44] Dapor, M. A monte carlo investigation of secondary electron emission from solid targets: Spherical symmetry versus momentum conservation within the classical binary collision model. *Nuclear Instruments and Methods in Physics Research Section B: Beam Interactions with Materials and Atoms* **267**, 3055 – 3058 (2009). URL http://www.sciencedirect.com/science/article/pii/S0168583X09007186. Proceedings of the Ninth International Conference on Computer Simulation of Radiation Effects in Solids.
[45] Rsler, M. & Brauer, W. Theory of secondary electron emission. ii. application to aluminum. *physica status solidi (b)* **104**, 575–587 (1981).
[46] Chang, H.-Y., Alvarado, A. & Marian, J. Calculation of secondary electron emission yields from low-energy electron deposition in tungsten surfaces. *Applied Surface Science* **450**, 190 – 199 (2018).
[47] Balcon, N., Payan, D., Belhaj, M., Tondu, T. & Inguimbert, V. Secondary electron emission on space materials: Evaluation of the total secondary electron yield from surface potential measurements. *IEEE Transactions on Plasma Science* **40**, 282–290 (2012).
[48] Ludwick, J. et al. Angular dependence of secondary electron yield from microporous gold surfaces. *Journal of Vacuum Science & Technology B* **38**, 054001 (2020).
[49] Gulans, A. et al. exciting: a full-potential all-electron package implementing density-functional theory and many-body perturbation theory. *Journal of Physics: Condensed Matter* **26**, 363202 (2014).
[50] Perdew, J. P. et al. Restoring the density-gradient expansion for exchange in solids and surfaces. *Phys. Rev. Lett.* **100**, 136406 (2008).
[51] Runge, E. & Gross, E. K. U. Density-functional theory for time-dependent systems. *Phys. Rev. Lett.* **52**, 997–1000 (1984).